\documentclass[usenatbib]{mn2e}
\usepackage{graphicx}
\usepackage{stfloats}
\usepackage{amsmath}
\usepackage{amssymb}

\title[Tidal Heating of Stars in the Galactic Center]{Accumulated
Tidal Heating of Stars Over Multiple Pericenter Passages Near SgrA*}

\author[Li and Loeb]{Gongjie Li, Abraham Loeb \\ Institute for Theory \& Computation, Harvard-Smithsonian CfA, 60 Garden Street, Cambridge, MA, USA}

\begin{document}
\topmargin-0.5cm
\bibliographystyle{mn2e}
\maketitle

\newcommand{\apj}{ApJ}
\newcommand{\apjl}{ApJL}
\newcommand{\apjs}{ApJS}
\newcommand{\mnras}{MNRAS}
\newcommand{\aap}{AAP}
\newcommand{\prd}{PRD}
\newcommand{\aj}{AJ}
\newcommand{\pasp}{PASP}
\newcommand{\araa}{ARA\&A}
\newcommand{\nat}{Nature}
\newcommand{\nar}{New Astron. Rev}
\newcommand{\apss}{Ap\&SS}
\newcommand{\actaa}{Acta Astron}
\newcommand{\be}{\begin{equation}}
\newcommand{\ee}{\end{equation}}
\newcommand{\bea}{\begin{eqnarray}}
\newcommand{\eea}{\end{eqnarray}}

\def\Mpc{\rm Mpc}
\def\Mbh{M_{\rm BH}}

\def\Msun{M_{\rm \odot}}
\def\kpc{\rm kpc}
\newcommand{\comment}[1]{}



\begin{abstract}

We consider the long-term tidal heating of a star by the supermassive
black hole at the Galactic center, SgrA*. We show that gravitational
interaction with background stars leads to a linear growth of the tidal
excitation energy with the number of pericenter passages near SgrA*. The
accumulated heat deposited by excitation of modes within the star over
many pericenter passages can lead to a runaway disruption of the star at a
pericenter distance that is 4-5 times farther than the standard tidal
disruption radius. The accumulated heating may explain the lack of massive
($\gtrsim 10M_{\odot}$) S-stars closer than several tens of AU from SgrA*.

\end{abstract}

\begin{keywords}
black hole physics -- galaxies: nuclei -- binaries:close -- stars: oscillations
\end{keywords}


\section{Introduction}
\label{s:intro}

Near the Galactic center, stars may get scattered into orbits for
which the tide raised by the supermassive black hole, SgrA*, at
pericenter is large but not strong enough to disrupt the stars. The
scattering rate into those orbits is larger than that of immediate
tidal disruptions orbits, where the pericenter distances are smaller
than the tidal radius, $r_p\lesssim r_t= R_* ({M_{\rm
BH}}/{M_*})^{\frac{1}{3}}$ \citep{Magorrian99, Alexander01}. Here
$M_{\rm BH}=4\times 10^6M_{\odot}$ is the mass of SgrA* \citep{Ghez08,
Genzel10}, and $M_*$ and $R_*$ are the mass and radius of the star.
In the near miss regime, stars with $r_p\gtrsim r_t$ are not disrupted during their
first passage near SgrA*, their tidal heating and bloating could still
be substantial after multiple passages due to the tidal distortion and
the excitation of internal oscillation modes. In principle, a
sufficiently large number of close passages may lead to the disruption
of these stars \citep{Rees88, Novikov92, Kosovichev92, Diener95,
Alexander03, Antonini11, Guillochon12}. Various tidal effects at $r_p\gtrsim r_t$
were considered in the literature, including relativistic effects \citep{Luminet85,
 Gomboc05, Ivanov06, Kosti09}, tidal heating of planets by stars \citep{Ivanov04_2, Ivanov07,
Ivanov11}, and tidal heating in close binary systems \citep{Press77,
Kochanek92, Mardling95_1, Mardling95_2, Lai97, Ho99, Ivanov04, Lai06,
Fuller11, Weinberg12}.
  
In this paper we consider the heating of stars at distances
$r_p\gtrsim 3 r_t$ from SgrA*. Since each pericenter passage is
associated with a small distortion in the shape of stars, one may
adopt a linear description for the tidal excitation of stellar modes
\citep{Novikov92, Kosovichev92}. The associated theory of linear mode
excitation has been calibrated recently by new data on stellar
binaries from the Kepler satellite \citep{Fuller12, Burkart12}. The
underlying theory was also recently extended to describe nonlinear
coupling of the excited modes \citep{Weinberg12}. We use the latest
results from these studies to calculate the tidal excitation and
heating of stars in the vicinity of SgrA*.

Our goal is to find the maximum distance from SgrA*
at which the accumulated heating due to numerous pericenter 
passages can lead to tidal disruption of stars around SgrA*.
The accumulated heating would lead to the absence of massive stars on eccentric orbits
interior to a spherical region around SgrA*, whose radius depends on
$M_\star$ and exceeds the standard tidal disruption radius $r_t$. Our
predictions could be tested by future searches for stars at closer
separations than the known S-stars, which have $r_p\gtrsim
10^2~{\rm AU}$ \citep{Ghez08, Genzel10}.

SgrA* is surrounded by a circumnuclear disk of young stars \citep{Genzel10}.
Inside the inner radius of this disk, there is the S-cluster of young
main sequence B-stars \citep{Ghez03, Eisenhauer05}, with random
orbital orientations and high orbital eccentricities
\citep{Gillessen09}. All the known S-stars have $r_p \gg r_t$, but it
is possible that the lack of S-stars inside 100 AU is caused by the
accumulated tidal heating over multiple pericenter passages.  Our
predictions can be tested as new stars, such as SO-102
\citep{Meyer12}, are being discovered and new instruments, such as
the second-generation VLTI instrument GRAVITY \citep{Bartko09}, are
being constructed.

The outline of the paper is as follows. In \textsection \ref{s:met} we
describe the method we use to calculate the heating due to tidal
excitation and the response of the stars. In \textsection \ref{s:res}
we show examples of these effects in the Galactic center using two
stellar models produced by MESA stellar evolution code
\citep{Paxton11} and present the results. In \textsection
\ref{s:conc}, we summarize our main conclusions.

\section{Heating of Stars by Tidal Excitation of Modes}
\label{s:met}

The tidal force from SgrA* can excite internal oscillation modes
within an orbiting star during its pericenter passages.  At distances
$r_p\gtrsim 3r_t$, the energy gain by tidal excitation per pericenter
passage is low, but the accumulated energy after many passages can
heat the star significantly.

\subsection{Mode Excitation and Interference in Multiple Pericenter Passages}

To calculate the low energy gain per orbit at $r_p\gtrsim 3 r_t$, it is appropriate to use
the linear perturbation formalism of \citet{Press77} \citep[see also][]{Novikov92,
Kosovichev92}. We denote the separation of the star from SgrA* at
time $t$ by $r(t)$. For a single passage, the energy of an excited
stellar mode can be expressed as,
\begin{equation}
\Delta E_{0,nml} = 2 \pi^2 \Big( \frac{G M_*^2}{R_*}\Big) \Big(\frac
{M_{\rm BH}}{M_*} \Big)^2 \Big(\frac{R_*}{r_p}\Big)^{2l+2}
|Q_{nl}|^2|K_{nlm}|^2 ,
\end{equation}
\label{e:E0n}
where $n$ is the mode order and $\{l,m\}$ are the two spherical harmonic
indices. The excited modes have $l>1$, $-l<m<l$, and we adopt the
convention in which $n<0$ for g-modes and $n>0$ for p-modes.  The
coefficient $K_{nlm}$ represents the coupling to the orbit,
\begin{equation}
K_{nlm} = \frac{W_{lm}}{2\pi} \int _{-\infty}^{\infty} dt \Big(
\frac{r_p}{r(t)}\Big)^{l+1} \rm{exp} \{i [ \omega_n t + m {\Phi
(t)} ] \} ,
\end{equation}
where $\omega_n$ is the mode frequency, $\Phi (t)$ is the true
anomaly, and $W_{lm} = (-1)^{(l+m)/2}[\frac{4 \pi}{(2
l+1)}(l-m)!(l+m)!] ^{1/2}/ [2^l\frac{(l-m)}{2}!\frac{(l+m)}{2}!]$.
The `tidal overlap integral' $Q_{nl}$ represents the coupling of the
tidal potential to a given mode,
\begin{equation}
Q_{nl} = \int_0^1 R^2 dR \rho(R) l R^{l-1} [\xi_{nl}^{\cal R} + (l+1)\xi_{nl}^{\cal S}] .
\end{equation}
where $\rho(R)$ is the stellar density profile as a function of radius
$R$.  $\xi (R) = [\xi_{nl}^{\cal R} (R) \hat{e}_R +
\xi_{nl}^{\cal S} (R) R \nabla] Y_{lm} (\theta, \phi) $ is the mode
eigenfunction, with $\xi_{nl}^{\cal R}$ being its radial component and
$\xi_{nl}^{\cal S}$ being its the poloidal component.  The total
energy transferred from the orbit to the star in a single passage is
\begin{equation}
\Delta E_{0} =  \sum_{nlm} \Delta E_{0,nml} ~~.
\label{e:E0}
\end{equation}

Next, we consider the evolution of the modes as a result of multiple
pericenter passages. If the dissipation timescale of the modes is
longer than the orbital period, the modes remain excited and interfere
with newly excited modes during subsequent
passages. \citet{Mardling95_1, Mardling95_2} considered this
problem numerically and found two orbital parameter regions.  In one
of them the energy exchange between the mode and the orbits is
quasi-periodic and the amplitudes of the modes remain small. In the
other region, chaotic behavior is exhibited. \citet{Ivanov04}
(hereafter IP04) further explored this stability boundary using a
proxy $\alpha$, which characterizes the change of the phase due to the
orbital period change, where the period change is caused by the energy
transferred to the modes. By mapping the mode amplitude and phase of a
particular passage to those values at an earlier passage, IP04 found
that when $\alpha$ is larger than a threshold value $\alpha_c$, there
is a secular increase of mode energy. $\alpha_c$ depends on the phase 
of the mode in the first passage.

For Galactic center stars with $r_p\gtrsim 3 r_t$ around
SgrA*, the change in orbital period per passage provided by the
exchange between tidal excitation energy and orbital energy is too
small to increase the mode amplitude. Below we show that gravitational
scattering on stars and compact objects in the Galactic center could
naturally lead to a drift in the orbital period that allows the
amplitude of the excited modes to increase stochastically. 

Similar to IP04, we introduce the two-dimensional vectors
$\mathbf{x}_i$ to characterize the amplitude $A_i$ and the phase $\psi_i$ of
the excited modes at the $i^{th}$ passage: 
\bea x_i^1 &=& A_i \cos
(\psi_i) , \nonumber \\ x_i^2 &=& A_i \sin (\psi_i) . \eea 
Because different stellar modes act independently in the linear regime, we
focus here on one mode with frequency $\omega_n$. For the $(i+1)^{th}$
passage, \be \mathbf{x}_{i+1} = \mathbf{\cal R}(\phi_i)[
\mathbf{x}_i+\mathbf{e}] ,
\label{e:iteration}
\ee where $\phi_i = \omega_n P_{orb, i}$ (with $P_{orb,i}$ being the
orbital period for the $i^{th}$ passage), $e^1 = 1$, $e^2 = 0$ and
$\mathbf{\cal R}$ is the rotation matrix.\\

\begin{figure}
\includegraphics[width=3.3in, height=2.7in]{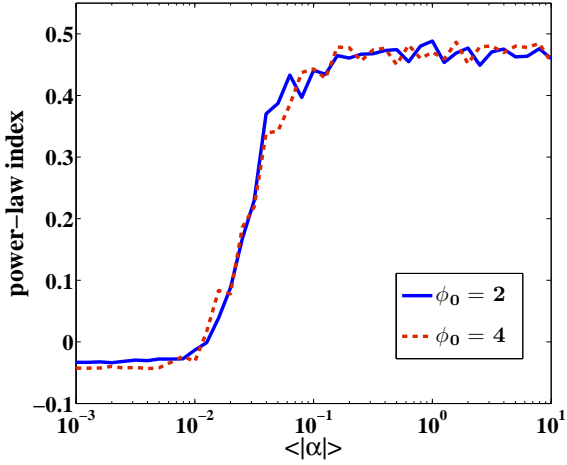}
\caption{\label{e:int} The power-law index of the mode amplitude
growth with time during multiple passages as a function of the average
magnitude of $\alpha = \omega_n \Delta P_{orb}$. When the power-law
index is around 0.5, the amplitude growth resembles a random walk and the energy of the mode is growing linearly with the number of
passages. We find this scaling when $\langle|\alpha|\rangle > 0.1$, independent of the
value of $\phi_0 = \omega_n P_{orb, 0} $ (shown by the different
lines).  }
\vspace{0.1cm}
\end{figure}

Defining $\alpha_i = \omega_n \Delta P_{orb, i}$ where $\Delta P_{orb,
i}$ is the change in the orbital period in the $i^{th}$ passage, we
get $\phi_{i+1} = \phi_i+\alpha_i$. In difference from IP04,
$\alpha_i$ is a random variable. Given the initial condition
$\mathbf{x}_0 = (1, 0)$ (without loss of generality) and equation
(\ref{e:iteration}), we examine numerically how the mode amplitude
changes as a function of the number of passages. First, we
examined the case when $\alpha$ is drawn from a uniform distribution
between $-2\alpha_m$ to $2\alpha_m$, $\langle |\alpha| \rangle =
\alpha_m$.  We characterize the growth in the mode amplitude by the
power-law index of its evolution with the number of passages (using a
total of $10^6$ passages). Figure \ref{e:int} shows that for
$\langle |\alpha| \rangle > 0.1$ the amplitude increases with a
power-law index of 0.5, so the energy of the mode increases linearly
with time. We also examined an alternative case with $\alpha$ drawn
from a Poisson distribution and the result was the same .

Note that in difference from IP04, the increase in the
amplitude is caused by the stochastic nature of $\alpha$.  We also
find that the threshold value does not show any dependence on
$\phi_0$. 

Next we examine the value of $\langle |\alpha| \rangle$ due to gravitational
perturbers in the Galactic center. We start by expressing $\alpha$ in
terms of the fractional change in the orbital period assuming the
primary excited mode has frequency $\omega_n \sim \sqrt{{N_{pm} G
M_*}/{R_*^3}}$ (with a typical value $N_{pm} \sim10$),  \bea \alpha
&=& \omega_{n} \Delta P_{orb} \nonumber \\
			&\sim& 3300 \frac{\Delta P_{orb}}{P_{orb}} \,
\Big[\sqrt{\frac{N_{pm}}{10}} \,\Big( \frac{1-0.9}{1-e}\Big)^{3/2}
\Big( \frac{r_{p}/r_t}{3}\Big)^{3/2}\Big] , \eea 
where $e$ is the orbital eccentricity.  Thus, when $\frac{|\Delta
P_{orb}|}{P_{orb}} \gtrsim 3\times10^{-5}$ the amplitude of the modes
increases stochastically.

We calculate the expected ${|\Delta P_{orb}|}/{P_{orb}}$ due to
gravitational scatterings using the N-body code BHINT
\citep{Lockmann08} to track the orbits of the stars and compact
objects in the Galactic center. We estimate ${|\Delta P_{orb}|}/{P_{orb}}$ 
for each passage, and the expectation value is calculated by 
averaging ${|\Delta P_{orb}|}/{P_{orb}}$ over $\sim50$ passages.
We performed a convergence test and
verified that our numerical errors are small and the results are
robust. The fractional change in the orbital period of a test star
depends on the semi-major axis $a$ and eccentricity $e$ of its orbit
and the distribution of perturbers within the S-cluster. We assume an
outer radius of $\sim 0.04$ pc ($=1''$) for the S-cluster, and
estimate the period change for typical S-stars with eccentricities in
the range of $0.85$--$0.95$. We consider the initial mass function
(IMF) that matches the mass distribution of S-stars inside $0.8''$
($dN/dm \propto m^{-2.15\pm 0.3}$) \citep{Bartko10}). The fractional
change of the orbital period is most sensitive to the massive stars
\citep{Murray-Clay12}. We normalize the IMF so that it gives $\sim 3$
S-stars with $M\sim20 M_{\odot}$ as observed. In the mass range of
$0.3$--$25 M_{\odot}$, the IMF yields a total of 800 stars.

We also considered the effects of scattering on stellar-mass black
holes (SBH) and a hypothetical intermediate-mass black hole (IMBH).
SBHs are more massive than the background stars and therefore are
expected to segregate in the Galactic center \citep{Morris93,
Miralda00, Freitag06}.  We normalize the number of SBHs (each having
$10M_{\odot}$) within 0.04 pc to be 1400, based on \citet{Miralda00}
and \citet{Freitag06}.  An IMBH was hypothesized as an agent for
randomizing the inclinations of stars in the S-cluster, potentially
creating the hyper-velocity stars and the stellar disk \citep{Yu03,
Sesana06, Yu07, Gualandris09, Perets10, Yu10}. To gauge its effect on
$\Delta P_{orb}$ we assume an IMBH mass of $10^3M_{\odot}$
\citep{Yu10} with either $a = 10^{-3}$pc ($= 206$AU)) and
$e = 0.80$ or $a = 3 \times 10^{-4}$pc and $e = 0.26$. The scattering due to the
SBH and IMBH dominate the fractional change in the orbital periods.

\begin{figure}
\includegraphics[width=3.3in, height=2.7in]{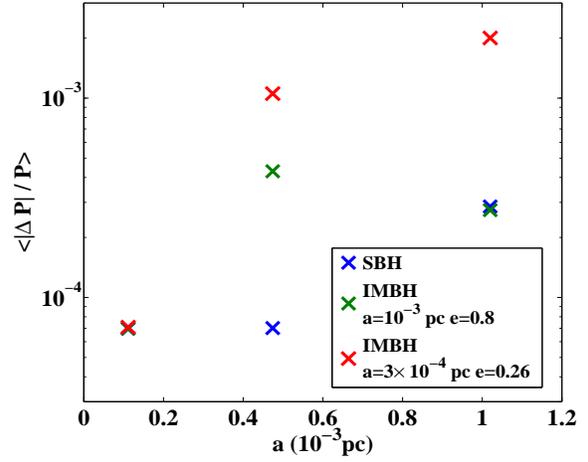}
\caption{\label{e:bhint} The average of the fractional change in orbital period per
pericenter passage, $\langle{|\Delta P_{orb}|}/{P_{orb}}\rangle$, for stars on orbits
with different semi-major axis $a$. The x-axis is in unit of $10^{-3}$
pc $ = 206$ AU. We include $800$ stars with an
initial mass function $(dN_*/dM_*) \propto M_*^{-2.15}$
\citep{Bartko10} in the mass range $0.3$--$25 M_{\odot}$, providing
about three $20 M_{\odot}$ stars. We also consider scattering on a
population of $1,400$ stellar-mass black holes (SBH) within $0.04$ pc
from Sgr A* \citep{Miralda00, Freitag06}, or an intermediate mass
black hole (IMBH) with a mass of $10^3M_\odot$ on two possible
orbits.}
\vspace{0.1cm}
\end{figure}

Figure \ref{e:bhint} shows the results from the numerical runs of the
N-body code.  We find that the fractional changes of the orbital
period are higher than the minimum value required to increase the mode
amplitudes stochastically, implying that the energy of the excited
modes would increase linearly with the number of pericenter passages.
Because the scattering of the orbit is fully random and the change of
the orbit is typically small ($\sim10^{-4}$--$10^{-3}$), we neglect
the orbital evolution. For a random walk, the period is
expected to change significantly only after $10^6$--$10^8$ passages,
beyond the number of passages considered here.

\subsection{Tidal Heating of Stars}
\label{s:HS}

Since the expected fractional change in the orbital period per
pericenter passage in Figure \ref{e:bhint} is higher than
$3\times10^{-5}$, the tidally-excited mode energy is expected to
increase linearly with the number of pericenter passages.
Cumulatively, a significant amount of heat might be deposited inside
the star during multiple passages. In this section, we consider the
dissipation of the mode energy and the resultant heating of the star.

Previous studies showed that when the amplitude of the excited modes
increases over some parametric instability threshold, the excited mode
begins to transfer its energy to lower frequency daughter modes
which dissipate rapidly \citep{Dziembowski82, Kumar96, Wu01, Arras03,
Weinberg08, Weinberg12}. We set $n_{crit} = E_{th}/\Delta E_0$ to be
the number of pericenter passages after which the amplitude of the
mode exceeds this threshold, where $E_{th}$ is the threshold energy
when non-linear coupling occurs. As the dissipation time of the
excited daughter modes is typically short compared with the
orbital period in the Galactic center, the thermal energy
gain in the stellar interior is:
\begin{equation}
E_{t, n_p} = (n_p/n_{crit}) E_{th} = n_p \Delta E_0 ,
\end{equation}
where $E_t$ is the thermal energy gained during this process, $n_p$ is
the number of pericenter passages and $\Delta E_0$ is the energy gain
of the excited modes during the first passage. When $n_p \gg
n_{crit}$, the thermal energy added to the star is independent of the
parametric instability threshold.

The heat generated around a radius $R$ within the star at time
$t_0$ will be trapped inside the star for a finite time,
$(t-t_0)<t_c(R)$, where $t_c(R)$ is the characteristic time it takes
heat to leak out. We estimate $t_c (R)$ as the minimum between the
photon diffusion time, $t_{diff}=\int dR
\{\tau(R)-(R_*-R)[d\tau(R)/dR]\}/c$, and the turbulent convection
time, $t_t=\int dR/v_c(R)$, for each spherical shell inside the star.
Here $\tau(R)$ is the scattering optical depth and $v_c(R)$ is the
convective velocity. At late times $t>t_c(R)$, the heating at radius
$R$ will saturate and reach a steady state where it is balanced
by cooling.  This sets the upper limit of the maximum heat stored at
a radius $R$.

As the non-linear coupling excites a large number ($>10^3$) of
daughter modes, most of the energy is redistributed. Typically, the
daughter modes consist of high order g-modes and so the energy is
redistributed mostly in the radiative zone. \citet{Weinberg12}
investigated modes inside solar-type stars and found that most of the
energy is transferred to the radiative core of the star. For
simplicity, we will assume that the energy is uniformly distributed
per unit mass within the radiative zone. 

The energy gained can be expressed as follows,
\be
 E_t(R) =
  \begin{cases}
   n_{crit}  \Delta E_0(R)       & \text{if } t_c (R) < P_{orb} n_{crit} \\
   \frac{t_c(R)}{P_{orb}} \Delta E_0(R)    & \text{if } t_c(R) > P_{orb} n_{crit} \\
  \end{cases}
\label{e:maxheat}
\ee Assuming that energy is evenly deposited throughout the entire
radiative zone of the star, we find $E_0(R)$ and obtain the thermal
energy stored at a radius $R$, $E_t(R)$. Typically for stars at
$r_p\gtrsim 3r_t$ around SgrA*, $n_{crit} \ll t_c(R) / P_{orb}$, and
so the total stored heat is independent of $n_{crit}$.  Finally,
integrating $E_t(r)$ over the interior of the star yields the total
heating inside the star, $E_H$.

As a result of the additional source of energy, the star expands. 
So far, we did not include the increase of
the size of the star in our calculation. As the stellar radius
increases, the tidal effects become stronger with $\Delta E \propto
R_*^6$. A decrease in the mode frequency $\sim \sqrt{{GM_*}/{R_*^3}}$
brings $\omega_n$ closer to the orbital frequency and increases
$K_{nl}$. Thus, ignoring the variation in the tidal overlap integral
($Q_{nl}$), the tidal excitation becomes stronger as the size of the
star increases. In addition, the rate of the expansion 
and the final size of the star depend on where the heat is deposited
\citep{Podsiadlowski96}. We examine this process more closely with MESA stellar
evolution simulations \citep{Paxton11} in the next section.

\begin{figure}
\includegraphics[width=3.3in, height=2.7in]{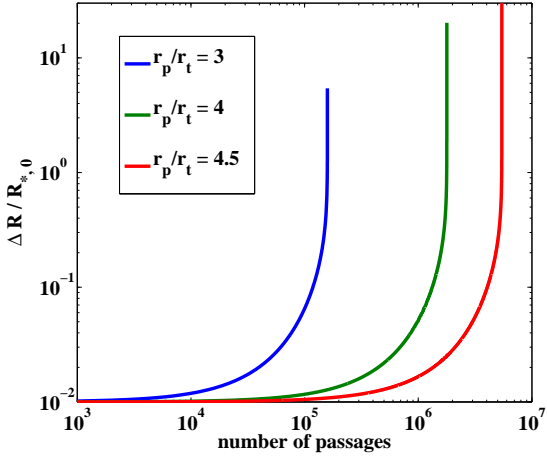}
\caption{\label{f:iterative} Radius of the star as a function of the
number of passages after saturation when $t>t_c(R=0)$, assuming
$\Delta R / R_{*, 0} (n = 0) = 0.01$. We find that a
star can be heated significantly after $\sim 10^6$ passages even if the
thermal energy it stores at saturation is only $1\%$ of its binding
energy. This threshold increases as $r_p/r_t$ increases.}
\vspace{0.1cm}
\end{figure}

As the star gains energy, its energy gain rate increases due to its
increasing size. The resulting runaway process could lead to the
disruption of the star. In order to find the minimum heating at
saturation ($t>t_c(R=0)$) that may lead to disruption, we express the
radius of the star after the $n^{th}$ pericenter passage as, $R_*(n) =
R_{*, 0} (1+\epsilon_n)$, where $R_{*, 0}$ is the original radius of
the star. Assuming $\Delta E \propto R_*(n)^6$ and ignoring the change
in entropy within the star, we find \be 1/(1 + \epsilon_n) - 1/(1 +
\epsilon_{n+1}) = \Delta \tilde{E_0}((1 + \epsilon_n)^6 - 1) , \ee
\label{e:iterative}
where $\Delta \tilde{E_0} = \Delta E_0 / ({GM_*^2}/{R_{*, 0}}) $.
Figure \ref{f:iterative} shows the growth of $R_*(n)$ as a function of
the number of pericenter passages starting with the saturation value
of $\epsilon_0 = 0.01$. Our
results demonstrate that at $r_p/r_t \sim$ 4 the stored heat can
approach the binding energy of the star after $\sim10^{6}$
pericenter passages following saturation, even if the total heat
gained at saturation is only $\sim1\%$ of the binding energy. This
threshold increases as $r_p/r_t$ increases.

During its lifetime, a massive star can achieve $\gtrsim 10^7$
pericenter passages at the corresponding distances from SgrA*. For
example, a 20 $M_{\odot}$ star with $e=0.9$ and $r_p\sim 5 r_t$ around
SgrA* has an orbital period of $\sim 0.8$ years. Thus, during its
lifetime the star encounters $\sim 10^7$ pericenter passages. The
maximum number of pericenter passages is also limited by gravitational
scatterings on other stars.  According to Figure \ref{e:bhint}, with
stochastic scatterings on SBH or one hypothesized IMBH, the maximum
number of passages at the original pericenter is $\sim
 10^6-10^8$. Thus, for $r_p/r_t < 5$, the star will be significantly heated
even if the total heat gained at saturation is only $\sim1\%$ of the
binding energy.

Non-linear effects are expected to dominate in the last phase of the
disruption process. When the star is distorted, the energy transfer
from the orbit to the modes can be either positive or negative
depending on the phases of the modes and the orientation of the
ellipsoid at the time of the pericenter passage. \citet{Diener95}
studied this effect statistically and found that the probability of a
positive transfer of energy from the orbit to the star is high.

\section{Results}
\label{s:res}
Based on the formalism presented in \textsection \ref{s:met}, we 
calculated the tidal heating of stars in the Galactic center.  We
consider two steller masses: $1 M_{\odot}$ (representing low-mass
stars) and $20M_{\odot}$ (representing high mass stars, similar to
SO-2 \citep{Martins08}). The other properties of the two stars are
summarized in Table \ref{t:model}.

\begin{table}
\caption{Properties of stellar models}
\begin{tabular}{|l||l|l|l|l|}
\hline
	Mass	&	Metallicity	 &	Radius 	&	Age	\\
   ($M_{\odot}$) &     & ($R_{\odot}$) & (yrs) \\
\hline
	1	&	$Z= Z_{\odot}$	&	1		&	$4.5\times10^9$	\\
	20	&	$Z= Z_{\odot}$	&	10	&	$7\times10^6$	\\
\hline
\end{tabular}
\medskip\\
\label{t:model}
\end{table}

Since the energy gain in each passage depends on
$\Big({R_*}/{r_p}\Big)^{2l+2}$, and because the value of $Q_{nl}$
and $K_{nlm}$ are similar for modes with different values of $l$, the
quadrupole ($l=2$) modes gain the most energy during the tidal
excitation (whereas $l=0$ and $l=1$ modes are not excited). Thus, we
focus on the $l=2$ modes. 

\begin{figure}
\includegraphics[width=3.3in, height=2.7in]{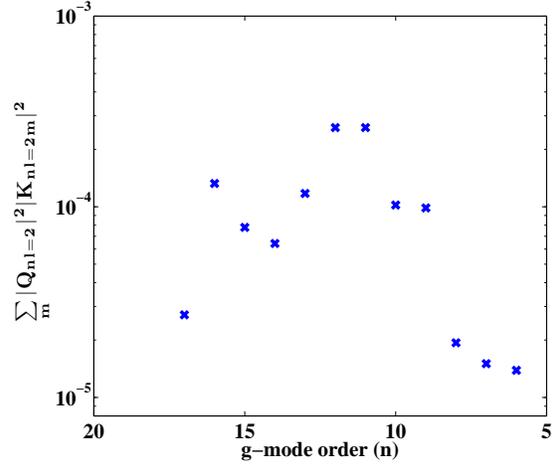}
\caption{\label{e:modes} $\sum\limits_m |Q_{n, l=2}|^2|K_{n, l=2, m}|^2$ as a
function of the mode order (n) for stellar modes computed with the
ADIPLS code \citep{Christensen07} based on the stellar structure from
the MESA stellar model \citep{Paxton11}. The mass ($20M_\odot$) and
radius ($10R_\odot$) of the star resemble those of SO-2
\citep{Martins08}. }
\vspace{0.1cm}
\end{figure}

We calculate the overlap integral ($Q_{nl}$) and the orbit coupling
($K_{nlm}$) using the MESA stellar model \citep{Paxton11}.  The
adiabatic normal modes are computed with the ADIPLS code
\citep{Christensen07}. For illustration, we show in Figure
\ref{e:modes} the values of $\sum\limits_m |Q_{n, l=2}|^2|K_{n, l=2, m}|^2$ for a
$20 M_{\odot}$ star in orbit around Sgr A* with $a =
7\times 10^{-3}$ pc and $e = 0.9$. As expected \citep[e.g.][]{Press77,
Burkart12}, we find that lower order g-modes are excited the most.  The
energy gain in one passage, $\Delta E_0$, can then be found from
equation (\ref{e:E0}).

\begin{figure}
\includegraphics[width=3.3in, height=2.7in]{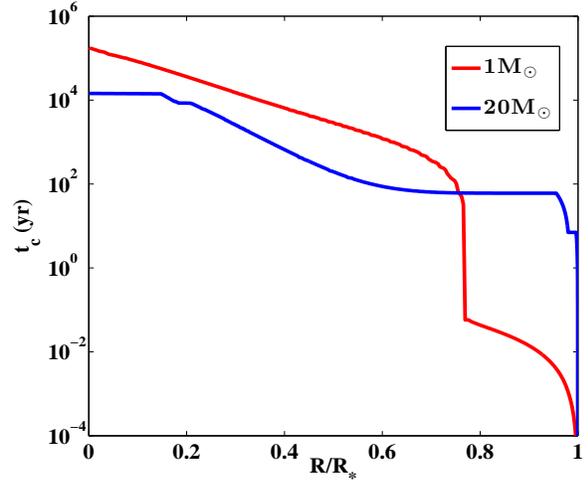}
\caption{\label{e:tcool} The cooling time ($t_c$) as a function of
radius for the two stars in Table \ref{t:model}.}
\vspace{0.1cm}
\end{figure}

To calculate the time it takes for the deposited heat to travel to the
surface ($t_c$) as described in \textsection \ref{s:HS}, we obtain the
optical depth and the convective velocity profile in the interior of
the stars from the MESA code \citep{Paxton11}. Figure \ref{e:tcool}
shows the cooling time as a function of radius for the two stars. 

The threshold for non-linear coupling has been discussed by
\citet{Weinberg12} for three mode coupling in a solar-mass star.  If
the daughter modes only couple to one other daughter mode, the
threshold is $E_{th} \sim 10^{-19} GM_*^2/R_*$; however, if the
daughter modes couple to multiple daughters, $E_{th} \sim 10^{-16}
GM_*^2/R_*$. In both cases, $(t_c/t_{orb}) \gg n_{crit}$ in the
interior of the stars for $r_p \gtrsim 3 r_t$. Thus, the heating of
the star is independent of the value of $n_{crit}$. Conservatively, we
calculate the heating using the high energy threshold.

\begin{figure}
\includegraphics[width=3.3in, height=2.7in]{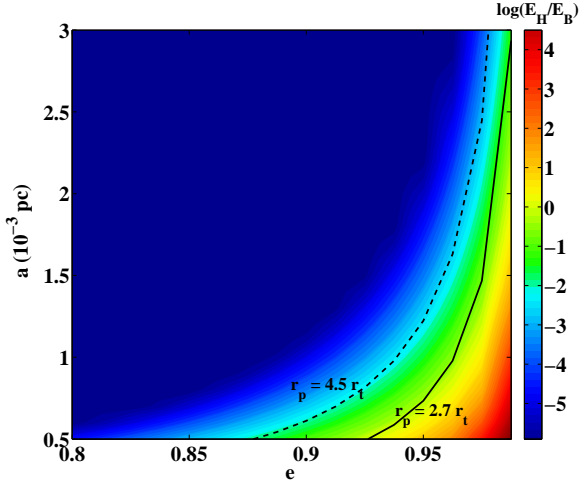}
\caption{\label{e:aeplot} Maximum amount of heat gained at saturation
in units of the binding energy of the star as a function of its
orbital parameters $a$ and $e$ for a $20 M_{\odot}$ star. The solid black line
indicates the pericenter distance boundary $r_p=2.7 r_t$ below which
the linear tidal excitation formalism is not applicable
\citep{Novikov92}. The dashed black line delineates the threshold
$E_H/E_B \sim 0.01$, beyond which the star can potentially be disrupted
in $\lesssim 10^6$ pericenter passages. We find that $20 M_{\odot}$ stars
are significantly heated at $r_p \sim 4.5 r_t$.}
\vspace{0.1cm}
\end{figure}

As the daughter modes consist of high order g-modes, energy is
redistributed mostly in the radiative zone of the star. For
simplicity, we assume that the distribution is uniform per unit mass
in the radiative zone and integrate equation (\ref{e:maxheat}) over
the interior of the star. Figure \ref{e:aeplot} shows the heat gained
by the stars ($E_H$) in units of their binding energy ($E_B$) obtained
from MESA, at the saturation time $t=t_c(R=0)$. The increase in the
stellar radius is not included in this calculation.

Taking account of the runaway increase in the stellar radius, the net
heat deposited could approach the binding energy and hence lead to
disruption when the heating at saturation approaches $ 1\%$ of the
binding energy. We estimate that the heating could be substantial at
$r_p \sim 4.5 r_t$ for $20 M_{\odot}$ stars.

Next we analyse the heating effect more accurately using MESA stellar
evolution simulations. We estimate the heating rate by $\Delta E_0 / P_{orb}$, and assume that the heat is deposited uniformly in
the radiative zone. We take account of the change in
$|Q_{nl=2}|^2|K_{nl=2m}|^2$ due to the change of the stellar structure
through iterations. For our first iteration, we assume a constant
$|Q_{nl=2}|^2|K_{nl=2m}|^2$ and obtain the structure of the heated
stars with different radii at different times. Then we calculate the
increase in $|Q_{nl=2}|^2|K_{nl=2m}|^2$ as a function of the increase
in stellar radius for the heated stars.  For our second iteration, we
simulate the heated stars with a changing $|Q_{nl=2}|^2|K_{nl=2m}|^2$
as a function of stellar radius. We calculate
$|Q_{nl=2}|^2|K_{nl=2m}|^2$ and continue iterating until the
dependence of $|Q_{nl=2}|^2|K_{nl=2m}|^2$ on radius converges. In the
examples we consider, convergence is reached within two iterations. 

\begin{figure}
\includegraphics[width=3.3in, height=2.7in]{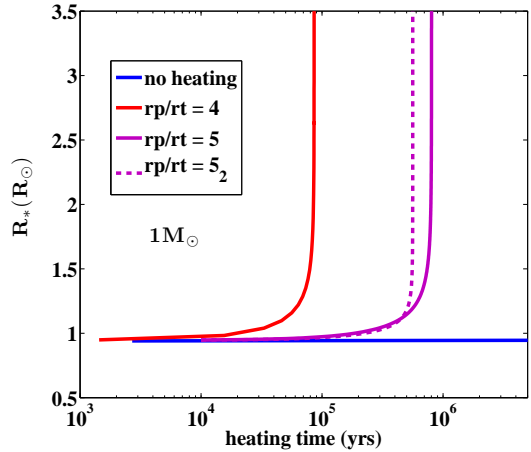}\\
\includegraphics[width=3.3in, height=2.7in]{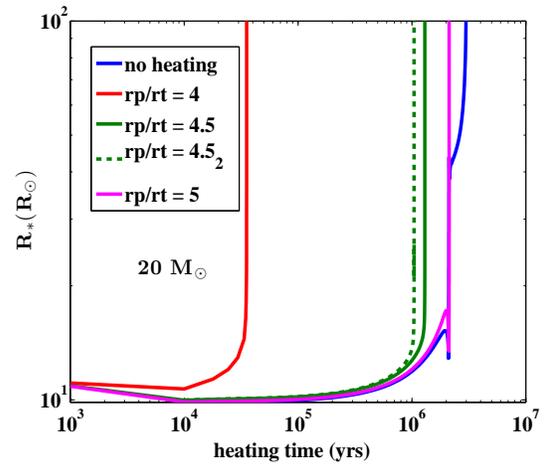}
\caption{\label{e:MESAheat} Stellar radius versus heating time. {\it
Top panel:} $1 M_{\odot}$ star; {\it Bottom panel:} $20 M_{\odot}$
star.  Blue lines indicate the radius change by stellar
evolution. Requiring the heating time to be shorter than the orbital
scattering timescale ($\sim 10^6$ yrs) and the lifetime of the
unheated stars, the maximum $r_p$ for which the stellar radius
significantly is $\sim 4.5 r_t$ for the 20 $M_{\odot}$ star, and $\sim
5 r_t$ for the 1 $M_{\odot}$ star.  At these limiting cases, the
dashed lines show results from a second iteration in which
$|Q_{nl=2}|^2|K_{nl=2m}|^2$ is updated as the stellar radius
increases.}
\vspace{0.1cm}
\end{figure}

Our convergent results for the 20 $M_{\odot}$ star indicate that the
size of the convective core decreases and the central temperature
stays approximately constant during the heating. For the 1 $M_{\odot}$
star, the size of the radiative core increases and the central
temperature drops significantly.  Figure \ref{e:MESAheat} shows the
radius of the heated star as a function of time.  We compare the
results of the two iterations for the 20 $M_{\odot}$ star at $r_p/r_t
= 4.5$ and for the 1 $M_{\odot}$ star at $r_p/r_t = 5$, and find that
the disruption time depends only weakly on the change in
$|Q_{nl=2}|^2|K_{nl=2m}|^2$. For other pericenter distances we show
only the results of the first iteration (assuming
$|Q_{nl=2}|^2|K_{nl=2m}|^2=$const).  Requiring the heating timescale
to be shorter than the orbital scattering timescale and the stellar
lifetime, we find that the maximum $r_p$ for disruption is $\sim 4.5
r_t$ for a 20 $M_{\odot}$ star and $\sim 5 r_t$ for a 1
$M_{\odot}$ star. 

For simplicity, we only considered non-rotating stars.  As discussed
by \citet{Fuller12}, the mode frequencies are modified for rotating
stars by $m C_{nl} \Omega_*$, where $\Omega_*$ is the rotation rate of
the star and $C_{nl} = \int^{R_*}_0 \, \rho R^2(2\xi^{\cal R}\xi^{\cal S}+\xi^{{\cal S}{2}})
\, dR$. Because $Q_{nl}$ are unchanged by rotation, the dominant modes
shift to higher order g-modes which have smaller values of
$Q_{nl}$. Thus, rotation would lower the excitation energies. In
addition, the rotation may modify the modes themselves
\citep{Burkart12}, and further complicate the calculation.  Treatment of
tidal excitation in misaligned spin-orbit systems were discussed by
\citet{Ho99} and \citet{Lai06}.

\begin{figure}
\includegraphics[width=3.5in, height=2.7in]{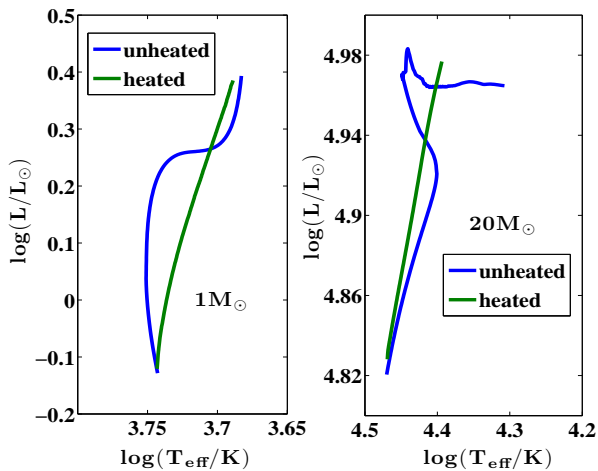}
\caption{\label{e:HR} HR diagram of heated stars with masses of $1
M_{\odot}$ (left panel) and $20 M_{\odot}$ (right panel). Blue lines
indicate the evolution track of giant stars with the same masses as
they evolve off the main sequence. The HR diagrams of the heated stars
stop at the point when the tidal radii of the heated stars approach
$(r_p/2.7)$, at which point the linear tidal excitation approach
breaks down.}
\vspace{0.1cm}
\end{figure}

Finally, we discuss the observational signature of a tidally heated
star.  Using the MESA simulation, we plot the Hertzsprung-Russell (HR)
diagram of the heated stars in Figure \ref{e:HR}. Because our
calculation is not appropriate in the non-linear regime when the tidal
radius of the heated star approaches $\sim (r_p/2.7)$, we stop the
calculation when $r_p \sim 2.7 r_{t, heated}$, where $r_{t, heated}$
is the tidal radius of the heated star. We find that a 1 $M_{\odot}$
star at $r_p \sim 2.7 r_{t, heated}$ acquires a luminosity $L$ that is
$\sim 3$ times higher than if it were on the main sequence and an
effective temperature $T_{eff}$ that is $\sim 12 \%$ lower than the
main sequence star.  A 20$M_{\odot}$ star at $r_p \sim 2.7 r_{t,
heated}$ acquires a luminosity that is $\sim 44 \%$  times higher and an
effective temperature that is $\sim 20 \%$ lower than that on the main
sequence.  Photometrically, the heated stars could be confused with
giant stars that evolved off the main sequence (illustrated by the
blue lines in the plot).

\section{Conclusions}
\label{s:conc}
We considered the tidal excitation of oscillation modes in stars
orbiting SgrA*. When the dissipation timescale of the modes is longer
than the orbital period, the modes excited in each passage
interfere. Due to the gravitational scatterings on nearby stars or
stellar-mass black holes, the orbital period of the excited star
changes stochastically and the energy of the excited modes increases
approximately linearly with the number of pericenter passages. As
non-linear coupling of the stellar modes dissipate the kinetic energy
of the modes, the excited star is heated. Once the deposited heat is
significant, the star bloats and its tidal heating accelerates, until
non-linearities lead to the final mass loss and possible disruption of
the star.

We calculated the thermal energy gain by a star as a function of the
semi-major axis and eccentricity of its orbit around Sgr A*. We have
found that the maximum pericenter distance where the heat gained by
the star approaches its binding energy is $r_p \sim 5 r_t$ ($\sim
3.7$ AU) for a 1 $M_{\odot}$ star and $r_p \sim 4.5 r_t$ ($\sim 13$ AU)
for a 20 $M_{\odot}$ star.  The accumulated heating may explain the
lack of massive ($\gtrsim 10M_{\odot}$) S-stars closer than several
tens of AU from SgrA* \citep{Genzel10}.

The heating process may be most effective for the highest-mass stars
($\gtrsim100 M_{\odot}$), where radiation pressure nearly balances
gravity and reduces the binding energy considerably relative to
${GM_*^2}/{R_*}$ \citep{Shapiro86}. This makes these stars more
vulnerable to disruption through heating. However, the heating is not
important for giant stars evolved off the main sequence, because for
$r_p/r_t \sim 5$ the orbital period of a giant star is too long to
allow sufficient number of pericenter passages during the star's
lifetime.

The expected radius of the cavity produced by tidal disruption of
stars depends on stellar mass \citep{Alexander01}. Since gravitational
scatterings on other objects could change the orbital period on a
timescale much shorter than the lifetime of a low mass star but
similar to the lifetime of the high mass star ($\sim 20 M_{\odot}$),
the net number of pericenter passages is similar in
the two cases. Of course, the tidal distance of a high mass star is
larger than that of a low mass star, and so a lower mass star may
approach Sgr A* at a closer distance (having a shorter orbital time
and more pericenter passages) before being tidally disrupted.

The removal of tidally-heated stars makes it more difficult to test
the no hair theorem of general relativity based on stellar orbits, as
the precession produced by the quadruple moment of SgrA* decreases
with increasing distance. For example, the precession rate due to the
quadruple moment of SgrA* is only $\sim 0.4 \rm{\mu as/yr}$ for a
$20M_{\odot}$ star with $r_p = 4.5 r_t$, and is $\sim 4 \rm{\mu as/yr}$
for a $1M_{\odot}$ star with $r_p = 5 r_t$, assuming a normalized
spin of 0.7 for SgrA* \citep{Will08}. Gravitational deflections by
other stars or compact objects contaminate the precession signal and
require the monitored stars to be within $\sim 2 \times 10^{-4}$ pc
from SgrA* \citep{Merritt10}. We find that only low mass stars (which
cannot be detected at present) would be viable targets for testing the
no hair theorem around SgrA*.

As new stars, such as SO102 \citep{Meyer12}, are being discovered in
the Galactic center, our predictions for the tidal cavity radius as a
function of stellar mass may be tested. In particular, the
second-generation VLTI instrument GRAVITY will be able to resolve
faint stars with a K-band magnitude $m_K = 18$ ($\sim 3 M_{\odot}$)
\citep{Bartko09} and test our predictions in the coming years.


\section*{Acknowledgments}
We thank Charlie Conroy, James Fuller, Bence Kocsis, Leo Meyer, Sterl Phinney, Eliot
Quataert and Nick Stone for helpful comments. GL benefitted 
significantly from the MESA 2012 summer school. This work was supported
in part by NSF grant AST-0907890 and NASA grants NNX08AL43G and
NNA09DB30A.

\bibliography{msref}

\end{document}